# Vector vorticity of skyrmionic texture: An internal degree of freedom tunable by magnetic field

Xiaoyan Yao* and Shuai Dong[†]

*School of Physics, Southeast University, Nanjing 211189, China*

[Abstract] Different from the skyrmion driven by the Dzyaloshinskii-Moriya interaction in non-centrosymmetric materials, the skyrmionic texture in centrosymmetric magnet may possess the extra internal degrees of freedom, which greatly enrich its morphologies, imply the continuous deformation allowed by topological protection, and enable flexible tunability and potential functionality. To describe the internal degree of freedom to full extent, the conventional integer-valued scalar vorticity is extended into a vector vorticity with continuous rotation allowed. The further simplified vorticity angle, together with helicity angle, represents the whole rotation freedom of spin space. The centrosymmetric magnet with frustration provides a perfect platform for realizing a controllable manipulation on these internal degrees of freedom, where the magnetic field can be applied to tune vorticity continuously in both crystal and isolated forms of skyrmionic texture, and the helicity can be further controlled by electric field. Moreover, the simulation reveals the distinctive dynamic effects related to the vorticity modulation, namely, a straight motion can be generated by rotating magnetic field to tune vorticity, and the vorticity can also be controlled to modulate the dynamics induced by the spin polarized current.

Corresponding author

*yaoxiaoyan@seu.edu.cn

[†]sdong@seu.edu.cn

## I. INTRODUCTION

In the past half century, topology had become an important part of condensed matter physics. Besides exotic topological states manifesting in reciprocal space [1,2], topological structures existing in real space as localized magnetic textures also attracted a great deal of interest from both the academic and technological fields [3-9]. Among them, magnetic skyrmion, a topologically stable whirling spin texture, came into highlight in recent years. Since the experimental discovery in 2009 [10], magnetic skyrmions have been observed in many different materials [11-20]. Their compelling features, such as stable particle nature, small size and electric controllability, generate enormous interest in the future spintronic and topological applications [21-31]. On the one hand, the fidelity of topological nature guaranteed by the topological protection promises great potential for applications to non-volatile information storage and processing devices; on the other hand, the continuous deformation allowed by topological protection enables internal degrees of freedom, which imply extra tunability and functionality by external stimuli with invariant topological property.

The traditional magnetic skyrmions are mostly observed in non-centrosymmetric materials or heterostructures, which are primarily induced by the Dzyaloshinskii-Moriya interaction (DMI) [32-37]. Since the DMI is determined by the spin-orbit coupling under inversion symmetry breaking, its dependence on underlying crystal structure usually locks the magnetic texture, and therefore no internal degree of freedom remains. Meanwhile, skyrmions were also observed in some centrosymmetric materials [18,38-42], and the mechanism could be the frustration of exchange interactions [43-48], the four-spin interactions [19], the long-ranged magnetic dipole interactions [18], or the coupling between itinerant electron spins and localized spins [49,50]. Different from the non-centrosymmetric systems, centrosymmetric structure allows the internal degrees of freedom to exist in the skyrmions. Recently, as the skyrmions in smaller size with stronger topological Hall effect were experimentally discovered in the centrosymmetric materials [51-54], the internal degrees of freedom, as the unique character of skyrmions in the centrosymmetric structure, are attracting growing attention in theoretical and experimental research.

For the perpendicularly magnetized skyrmions, the internal degrees of freedom are described by helicity and vorticity [44-48]. Helicity denotes the angle between the in-plane component of the magnetization and the radial direction. For the dipole-dipole interaction induced skyrmion, two helicity values are allowed with degenerate energy [18]. For the skyrmion driven by frustration, the helicity is entirely unlocked and the continuous value from $-\pi$ to $\pi$ is allowed [46]. Different methods have been discussed to control the helicity [55-59], and our previous work predicated the tunability of helicity by applying an external electric field [58]. In contrast to the continuous helicity, the conventional vorticity shows discrete integer values, e.g. 1 and -1 for skyrmion and anti-skyrmions. (Here, skyrmion and anti-skyrmion are differentiated by vorticity following Ref. [4].) However, it is noteworthy that skyrmion and anti-skyrmion may have the same topological property, i.e. skyrmion, bimeron, and anti-skyrmion may be related by a rotation of each spin around an in-plane axis [60-62]. This implies a continuous variation of vorticity allowed by topological protection, and it is prospective to be realized in centrosymmetric materials. Since the conventional scalar definition of vorticity is not sufficient to describe this continuous degree of freedom, the corresponding extension and the possible manipulation in the centrosymmetric system become an essential and urgent problem.

In this article, the vorticity is extended from the conventional scalar to a vector to embody the continuous variation under topological protection. The rotation freedom of spin space can be further represented by the simplified vorticity and helicity angles. Thus a complete description is presented to the internal degrees of freedom for skyrmion in the centrosymmetric system. The continuous variation of vorticity produces rich morphologies from skyrmion to anti-skyrmion via multiple-meron state, which are generally called skyrmionic texture (SK) for convenience. The frustrated centrosymmetric system provides a perfect platform to realize the tunability of these internal degrees of freedom. The simulation indicates that the vorticity can be effectively modulated by magnetic field for both crystal and isolated forms of SK, and electric field can be further applied to control helicity. Even though the SK will be distorted by uniaxial anisotropy, the topology can be conserved and the effective modulation can be realized as long as the anisotropy is not too strong. As a degree of freedom, the vorticity also affects the SK dynamics, and thus it can be rotated by magnetic field to generate a straight shift, or modulated to control the motion driven by spin-polarized current.

The remainder of this paper is organized as follows. In Sec. II, based on the discussion about the topological theory, the vorticity is generalized to a vector form and then simplified to the vorticity angle. In Sec. III, the manipulation of vector vorticity and helicity is studied by simulation. In Sec. IV, the distinctive dynamic effects related to the vector vorticity are explored.

II. EXTENDED VORTICITY

According to the topological theory, the magnetic texture can be classified by a topological invariant, i.e. winding number, which serves as a fingerprint for a topological equivalence class [63]. The magnetic texture is topologically protected against continuous deformation into another texture with different winding number. For the isotropic three-dimensional (3D) spins, the spin space is 2-sphere $S^2$, the second homotopy group $\Pi_2(S^2)=Z$, where integer Z can be calculated by the $S^2$-winding number, i.e. topological charge or skyrmion number ($N_{sk}$) in the following form,

$$N_{sk} = \frac{1}{4\pi} \iint \mathbf{S} \cdot (\partial_x \mathbf{S} \times \partial_y \mathbf{S}) \, dxdy \,. \tag{1}$$

where $\mathbf{S}$ represents the spin vector normalized to 1. $N_{sk}$ quantifies the number of times spin vectors wrapping around a unit sphere as the coordinate ($x$, $y$) spans the defined region, which can be calculated for a spin lattice in the manner as Ref. [64]. Here SK means nonzero $N_{sk}$. To elucidate the detailed topological structure, the local topological charge density $\rho(\mathbf{r})$ is evaluated as following,

$$\rho(\mathbf{r}) = \frac{1}{4\pi} \mathbf{S} \cdot (\partial_x \mathbf{S} \times \partial_y \mathbf{S}) \,. \tag{2}$$

If the polar coordinates ($r$, $\varphi$) are introduced in the lattice plane, and the spin vector is represented by the usual spherical coordinates as $\mathbf{S}=(\sin\theta\cos\phi, \sin\theta\sin\phi, \cos\theta)$, then

$$N_{sk} = \frac{1}{4\pi} \iint \sin\theta (\partial_r \theta \partial_\varphi \phi - \partial_\varphi \theta \partial_r \phi) \, drd\varphi \tag{3}$$

Considering a conventional SK magnetized perpendicular to the lattice plane, using its circular symmetry with the center set as origin, $\theta$ only depends on $r$, i.e., $\theta(r)$, and $\phi$ only depends on $\varphi$, i.e., $\phi=v\varphi+\eta$, where $v$ is the conventional scalar vorticity and $\eta$ is the helicity. Then

$$N_{sk} = \frac{1}{4\pi} \iint \sin\theta \partial_r \theta \partial_\varphi \phi \, drd\varphi = \frac{1}{2} \int_0^\infty \sin\theta \partial_r \theta \, dr \cdot \frac{1}{2\pi} \int_0^{2\pi} \partial_\varphi \phi \, d\varphi = P_m v \,. \tag{4}$$

Here $P_m$ is polarity, giving the orientation of the spin in the core.

Note that when the spin space is $S^1$ for the strict easy-plane spins, the first homotopy group $\Pi_1(S^1)=Z$, and $v$ is just the winding number ($S^1$-winding number), which counts how many times the spins rotate around the unit circle when following a closed curve. Vortices ($v=1$) and anti-vortices ($v=-1$) can be distinguished topologically. When the spin space expands from easy-plane $S^1$ to spherical $S^2$, spins are not limited within the plane anymore. $v$ can be calculated by accumulating the angle variation of the in-plane spin components along the contour. Skyrmion ($v=1$) and anti-skyrmion ($v=-1$) can also be distinguished by $v$, but $v$ is not topological fingerprint anymore owing to $\Pi_1(S^2)=0$. The skyrmion number $N_{sk}$ takes its place as $S^2$-winding number. Meanwhile, the additional dimension of spin space provides a path for spin texture to escape the limitation of $S^1$-winding number [63], and thus the vorticity is endowed with a freedom to be tuned continuously under topological protection, which lays the theoretical foundation for the generalization of vorticity.

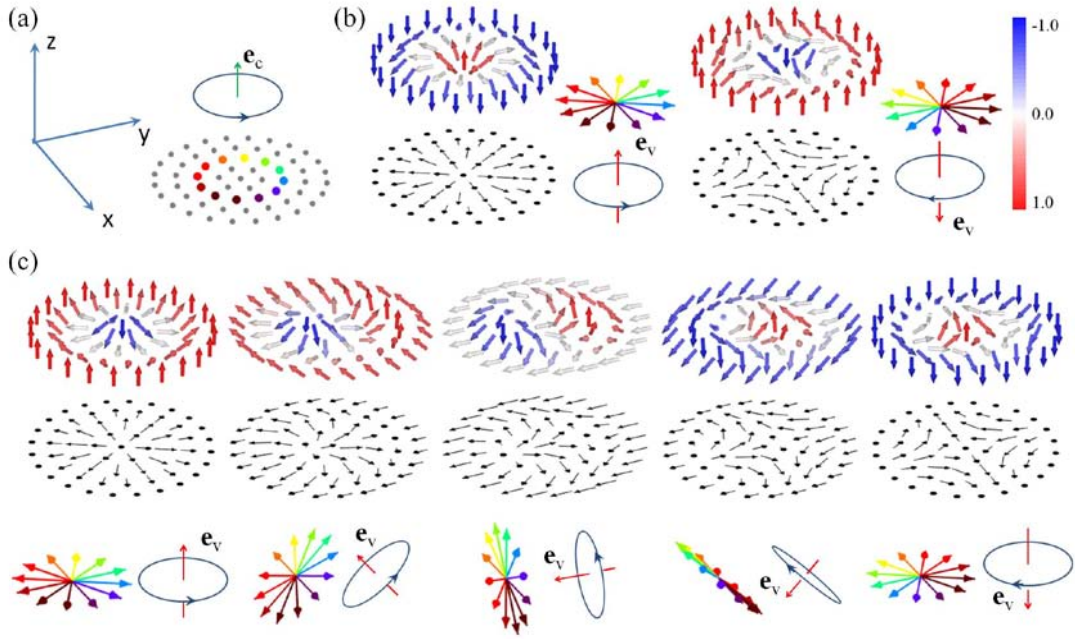

Figure 1 The schematic spin textures of SKs with $|N_{sk}|=1$. (a) Spin lattice lies in the $xy$-plane, on which a counterclockwise contour around SK core is marked by the colorful dots, and its orientation ($e_c$) is illustrated by green arrow upward along $+z$ direction. (b) Spin textures of $N_{sk}=1$ with $e_v$ (red arrow) along $+z$ on the left and along $-z$ on the right. (c) Spin textures of $N_{sk}=-1$ with $e_v$ rotating around $+x$ direction from $+z$ to $-z$ with the angle of 0, $0.25\pi$, $0.5\pi$, $0.75\pi$, and $\pi$ from left to right. The top row displays 3D vector maps of spins with the color referring to the $z$-components as shown in the right color bar. The corresponding projections onto $xy$-plane are presented below. The bottom row illustrates the spins on the contour with the color corresponding to (a), and then the orientation of $e_v$ (red arrow) is determined according to the spin curling direction by right hand rule.

Vorticity characterizes the rotation of spins following a contour on the crystal plane. This contour (C) could be assumed counterclockwise with the SK core as the center, and its orientation (unit vector $e_c$) can be assigned by the right hand rule, as illustrated by the arrow along $+z$ direction in Fig. 1(a). For the perpendicularly magnetized SK, the scalar $v$ is calculated by accumulating the angle variation of the in-plane spin components along the contour. Here the spin rotation plane just coincides with the crystal plane, as demonstrated in Fig. 1(b). When the spin space expands from easy-plane $S^1$ to spherical $S^2$, the spin rotation plane and the crystal plane may deviate from each other under topological protection. To describe this internal degree of freedom induced by additional dimension of spin space, vorticity can be generalized to a vector $V=Ve_V$, where $V$ presents the magnitude and $e_V$ shows

the orientation, namely

$$V = \left| \frac{1}{2\pi} \oint_c \partial_l \phi \, dl \right| \tag{5}$$

$$\boldsymbol{e}_V = \frac{\oint_c \boldsymbol{S} \times \partial_l \boldsymbol{S} \, dl}{\left| \oint_c \boldsymbol{S} \times \partial_l \boldsymbol{S} \, dl \right|} \tag{6}$$

Here, the variation of $\phi$ is accumulated for spin components within the rotation plane along the contour C. The normalized vector $\boldsymbol{e}_V=(e_{Vx}, e_{Vy}, e_{Vz})$ represents the normal direction of spin rotation plane, which is determined by the spin curling along the contour C for a SK. When the SK transforms continuously under topological protection, $e_v$ rotates continuously, and thus the vector vorticity $\boldsymbol{V}$ changes continuously.

In the conventional case of perpendicular magnetization, $\boldsymbol{V}$ is perpendicular to the lattice plane, then $v=\boldsymbol{V}\cdot\boldsymbol{e}_c$. Skyrmion appears as $\boldsymbol{e}_V$ is parallel to $\boldsymbol{e}_c$, and anti-skyrmion emerges with $\boldsymbol{e}_v$ antiparallel to $\boldsymbol{e}_c$, as demonstrated in Fig. 1(b). Otherwise the variation of $\boldsymbol{V}$ produces diverse morphologies under topological protection. As plotted in Fig. 1(c) with $N_{sk}=-1$ for instance, as $\boldsymbol{e}_V$ rotates from $+z$ to $-z$ direction, spin texture changes continuously from skyrmion to anti-skyrmion via the intermediate SKs during the process. In particular, the bimeron state appears when $\boldsymbol{e}_V$ just lies in $xy$-plane, which attracted considerable interest very recently [65-69].

In the perpendicularly magnetized case, the conventional skyrmion and anti-skyrmion show very different symmetry. The skyrmion with $v=1$ is very special for its axisymmetric spin textures of all helicities, and therefore no angle-dependent phenomena occurs. In contrast, anti-skyrmion with $v=-1$ possesses two-fold rotational symmetry with respect to the core as shown in Fig. 1. The strong anisotropic shape produces strongly angle-dependent (helicity-dependent) phenomena [70]. The continuous transformation from skyrmion to anti-skyrmion can be realized by rotating $\boldsymbol{V}$, namely, $\boldsymbol{V}$ can be tuned to switch between the different symmetries with $N_{sk}$ conserved.

It should be mentioned that from the same perpendicularly magnetized initial state, the rotation of $\boldsymbol{V}$ around different axis produces different SKs (See Note 1 of Supplemental Material for details [71]). This difference can be ascribed to the variation of helicity, which means that helicity and vorticity are entangled to some extent. To remove the overlap, it is reasonable to consider the rotation around $+x$ direction as a standard. Then the vector $\boldsymbol{V}$ can be reduced to the vorticity angle ($\lambda$), namely the angle of $\boldsymbol{V}$ from $+z$ direction. Thus helicity and vorticity can be described by two independent angle parameters: $\eta$ and $\lambda$, corresponding to the whole rotation freedom of spin space. Fig. 2 displays typical SKs with different ($\eta$, $\lambda$) schematically. In the isotropic system, these SKs with different ($\eta$, $\lambda$) are degenerate in energy.

Despite the topological equivalence between them, their magnetic and electric properties are distinct.

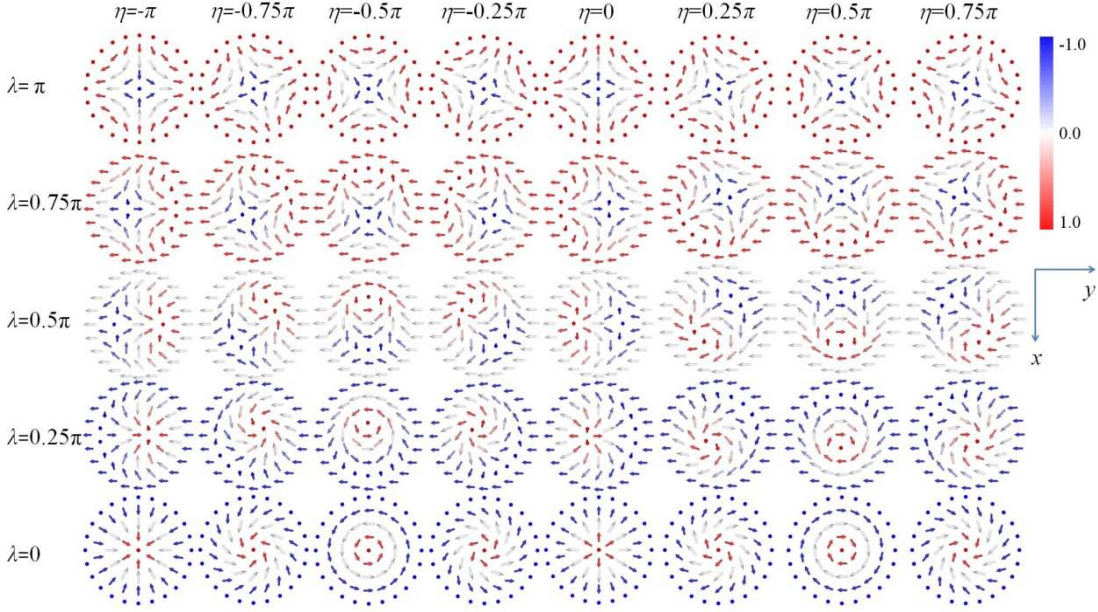

Figure 2 Top views of schematic SKs with helicity and vorticity degrees of freedom. Spin textures of $N_{sk}=1$ with $\lambda=0$, $0.25\pi$, $0.5\pi$, $0.75\pi$, and $\pi$ from bottom to top, and $\eta=-\pi$, $-0.75\pi$, $-0.5\pi$, $-0.25\pi$, $0$, $0.25\pi$, $0.5\pi$, and $0.75\pi$ from left to right. The color refers to the $z$-component of spins.

## III. MODULATION ON INTERNAL DEGREES OF FREEDOM

When the vector $V$ is considered, Eq. 4 can be rewritten as $N_{sk}=\boldsymbol{P_m}\cdot\boldsymbol{V}$, where the polarity $P_m$ is also extended to a unit vector $\boldsymbol{P_m}$, showing the core spin direction. Based on this equation, two topological modulations have been proposed. First, $N_{sk}$ can vary with $\boldsymbol{P_m}$, which is easy to be realized in the system with $V$ fixed by DMI. When the magnetic field ($\boldsymbol{h}$) is applied in the opposite direction normal to the plane, $N_{sk}$ reverses its sign [26,72]. Second, $N_{sk}$ may alter with $V$ when $\boldsymbol{P_m}$ is kept unchanged, which had been realized experimentally in the system with anisotropic DMI [73,74]. It is worth noting that there is the third variation remaining unsolved, i.e., as $N_{sk}$ is conserved, $V$ varies with $\boldsymbol{P_m}$ under topological protection, namely, $V$ can be controlled by $\boldsymbol{h}$, which is expected to be realized in the centrosymmetric system.

For the frustration-induced SK in centrosymmetric structure, the pure frustration origin keeps isotropic spin space, and thus endows SK with full internal degrees of freedom. Due to the topological character of SK, the spins on perimeter are antiparallel to the center one. Thus the magnetization $\boldsymbol{M}=M\boldsymbol{e_M}$, where $M$ denotes the magnitude and $\boldsymbol{e_M}=(e_{Mx}, e_{My}, e_{Mz})$ presents its orientation, will be antiparallel to $\boldsymbol{P_m}$, namely $N_{sk}=-\boldsymbol{e_M}\cdot\boldsymbol{V}$. Therefore, the manipulation of $V$ can be realized by applying ***h*** to control ***M*** under topological protection. To investigate the magnetic and electric behaviors related to these internal degrees of freedom, the atomistic spin dynamic simulation is performed on the two-dimensional frustrated triangular lattice with the Hamiltonian expressed as

$$H = -J_1 \sum_{<i,j>} \mathbf{S}_i \cdot \mathbf{S}_j - J_3 \sum_{<<i,k>>} \mathbf{S}_i \cdot \mathbf{S}_k - \mathbf{h} \cdot \sum_i \mathbf{S}_i - k \sum_i S_i^{z\,2} \qquad (7)$$

where $S_i$ represents a classic spin of unit length at the $i$-th site on $xy$-plane. The first and second terms on the right of Eq. 7 are the exchange energies. Ferromagnetic interaction $J_1$ between the nearest-neighboring spins is fixed as the energy unit, and all the parameters are simplified with reduced units. Antiferromagnetic interaction $J_3$ between the third-nearest-neighboring sites is considered to introduce frustration, which is similar to the case with the antiferromagnetic next nearest-neighboring interaction considered [43,75]. The third term describes the energy of the magnetic field (***h***=$h$***e***$_h$ with the magnitude $h$ and unit direction vector ***e***$_h$=($e_{hx}$, $e_{hy}$, $e_{hz}$)). And the fourth term represents the uniaxial anisotropy along $z$-axis with the magnitude $k$, which helps to generate the stable skyrmion lattice [45,46]. The simulation is performed on the triangular lattice of sites $N$=5184 with periodic boundary conditions. The system is relaxed by using the fourth-order Runge-Kutta method to numerically solve Landau-Lifshitz-Gilbert (LLG) equation as below.

$$\frac{\partial \mathbf{S}}{\partial t} = -\mathbf{S} \times \mathbf{h}_{eff} + \alpha \mathbf{S} \times \frac{\partial \mathbf{S}}{\partial t} \qquad (8)$$

where the Gilbert damping coefficient $\alpha$=0.2 is set to ensure quick relaxation to the equilibrium state, unless otherwise noted. $\mathbf{h}_{eff} = -\frac{\partial H}{\partial \mathbf{S}}$ is the effective field with Hamiltonian $H$ defined in Eq. 7. The time $t$ is measured in units of $\hbar/J_1$, and the lattice constant is adopted as the unit of length.

The SKs in the frustrated system may exist as isolated metastable excitations in the ferromagnetic background, or crystallize into stable lattice with nonzero anisotropy [45,46]. In the present article, isolated excitations are obtained at $J_3$=-0.3, $h$=0.1, and SK crystal in the case of $J_3$=-0.5, $h$=0.3. Both them survive with anisotropy switched off ($k$=0) owing to topological protection. When ***h*** is rotated around +$x$ direction from +$z$ to –$z$, ***M*** follows ***h*** strictly, and ***V*** is always parallel to ***M*** for $N_{sk}$=-1 (Fig. 3(a)) or antiparallel to ***M*** for $N_{sk}$=1 (Fig. 3(b)). Here $N_{sk}$ is evaluated for a single SK or each SK in crystal. Corresponding to Fig. 3(a), the variation of spin textures for SK crystal, namely from skyrmion crystal to anti-skyrmion crystal via bimeron crystal, is displayed in Figs. 3(c)-3(e). At the same time, the topological structure, namely the map of $\rho(\mathbf{r})$ keeps unchanged during the whole procedure of ***h*** rotation, as plotted in Fig. 3(f). Corresponding to Fig. 3(b), the isolated SK also shows the similar variation and keeps the same $\rho(r)$ map during the $h$ rotation as presented in Figs. 3(g)-3(j). Here, the simulation in this isotropic system confirms that $V$ can be strictly controlled by rotating $h$ for both crystal and isolated SKs with topological property unchanged.

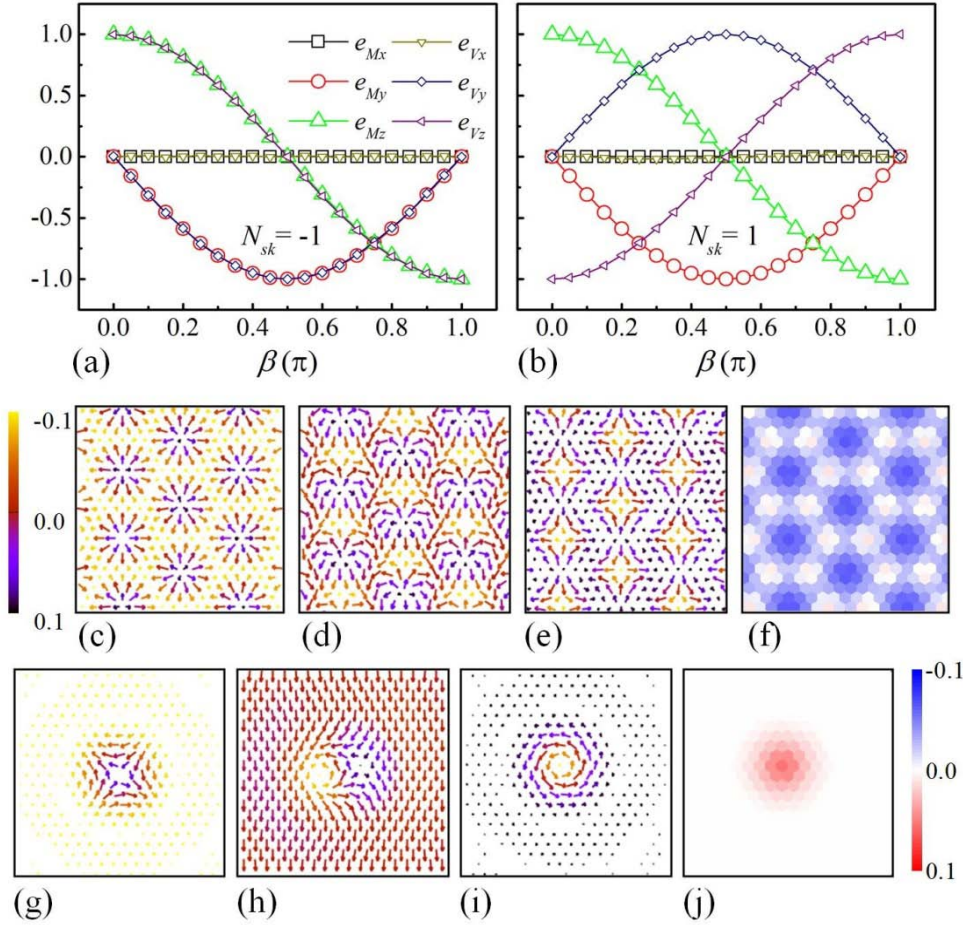

Figure 3 The simulated results for SKs with $h$ rotating around $+x$ direction from $+z$ to $-z$ in the isotropic system of $k=0$. The components of $e_M$ ($e_{Mx}$, $e_{My}$, $e_{Mz}$) and $e_V$ ($e_{Vx}$, $e_{Vy}$, $e_{Vz}$) as functions of $\beta$ ($\beta$ is the angle of $h$ from $+z$ direction) are plotted for (a) the crystal of SK with $N_{sk}=-1$, and (b) the isolated SK of $N_{sk}=1$. Accordingly, for the crystal of SK ($N_{sk}=-1$), the spin textures at (c) $\beta=0$, (d) $\beta=0.5\pi$, (e) $\beta=\pi$ and (f) $\rho(r)$ map are plotted in the middle. For the isolated SK of $N_{sk}=1$, the spin textures at (g) $\beta=0$, (h) $\beta=0.5\pi$, (i) $\beta=\pi$ and (j) $\rho(r)$ map are plotted below. The vectors in spin configurations show the projections of spins onto $xy$-plane, and the color refers to the $z$ components of spins as shown in the left color bar. The color bar on the right represents the value of $\rho(r)$. For visibility, only part of the lattice is plotted.

When the uniaxial anisotropy with $|k|=0.1$ normal to the lattice plane is switched on for the SK crystal, as illustrated in Fig. 4, $M$ deviates from $h$ with opposite tendencies for $k>0$ (Fig. 4(a)) and $k<0$ (Fig. 4(b)). But $V$ still follows $M$ closely, which ensures $N_{sk}$ well conserved during the rotation of $h$. Hence for the SK crystal $V$ can still be controlled effectively by $h$ as long as the anisotropy is not very strong. However, when $h$ tilts from the anisotropy direction, the topological map is distorted by the uniaxial anisotropy, and the distortion appears in different styles for $k>0$ and $k<0$, as discussed in Note 2 of Supplemental Material [71]. If the anisotropy is very strong, for instance $k>0.4$ for the SK crystal, the distortion may destroy the topological structure, and thus $V$ can not be smoothly controlled by $h$ any more. Similarly, the $h$ modulation of $V$ can also be realized for the isolated SK with uniaxial anisotropy, but the upper limit of $k$ for smoothly controlling $V$ is much lower than that of the SK crystal.

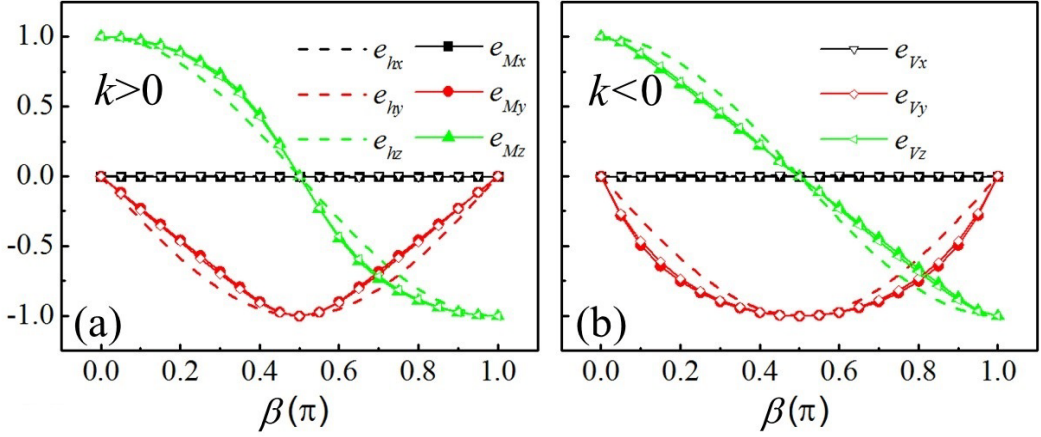

Figure 4 The simulated results for the SK crystal ($N_{sk}=-1$) with uniaxial anisotropy, as $h$ is rotated around $+x$ from $+z$ to $-z$ direction. The components of $e_h$ ($e_{hx}$, $e_{hy}$, $e_{hz}$), $e_M$ ($e_{Mx}$, $e_{My}$, $e_{Mz}$) and $e_V$ ($e_{Vx}$, $e_{Vy}$, $e_{Vz}$) as functions of $\beta$ ($\beta$ is the angle of $h$ from $+z$ direction) are plotted for (a) the easy-axis anisotropy $k=0.1$ and (b) the easy-plane anisotropy $k=-0.1$.

It should be mentioned that the noncollinear spin texture in the scale of the lattice constant usually produces the electric dipole $p_{ij} \propto -e_{ij} \times (S_i \times S_j)$ [76], where $e_{ij}$ denoting the unit vector connecting the two sites of neighboring $S_i$ and $S_j$. The total electric polarization ($P$) is estimated by the sum over all the bonds or all the local electric polarization on site ($p_i$). In the isotropic system with $k=0$, the SKs of $|N_{sk}|=1$ with different $\lambda$ exhibit different textures in $p_i$ map, as illustrated in Fig. S3(a) of Supplemental Material [71], resulting in $P$ with different magnitude and orientation. Fig. 5(a) displays 3D chart of $P$ dependent on $\eta$ with $\lambda$ varying from 0 to $\pi$ for the SK crystal of $N_{sk}=-1$. It is seen that the SK of a certain $\eta$ generates $P$ varying with $\lambda$ within a plane normal to $xy$-plane. Accordingly, Fig. 5(c) plots the magnitude of $P$ ($|P|$) as function of $\lambda$ for different $\eta$. It is seen that $P=0$ only for $\lambda=\pi$, or $\lambda=0$ and $\eta=\pm\pi/2$. Otherwise, $P$ shows nonzero value and different orientations. For a certain $\lambda$, the three components of $P$ show trigonometric functions of $\eta$, namely $P_x=P_{xm}(-\sin\eta)$, $P_y=P_{ym}(-\cos\eta)$, $P_z=P_{zm}\cos\eta$ as illustrated in the inset of Fig. 5(b). The amplitudes of $P$ components ($P_{xm}$, $P_{ym}$ and $P_{zm}$) depend on $\lambda$, namely $P_{xm}=P_{ym}\propto\sin\lambda$ and $P_{zm}\propto\cos\lambda+1$, as plotted in Fig. 5(b). Then $P_x\propto\sin\lambda(-\sin\eta)$, $P_y\propto\sin\lambda(-\cos\eta)$ and $P_z\propto(\cos\lambda+1)\cos\eta$ can be obtained. The distinction of **P** for different ($\eta$, $\lambda$) provides a base to tune helicity precisely on the condition of controlling vorticity in the isotropic system, that is, the precise manipulation of vorticity and helicity can be realized by applying magnetic field and electric field with appropriate orientation and magnitude.

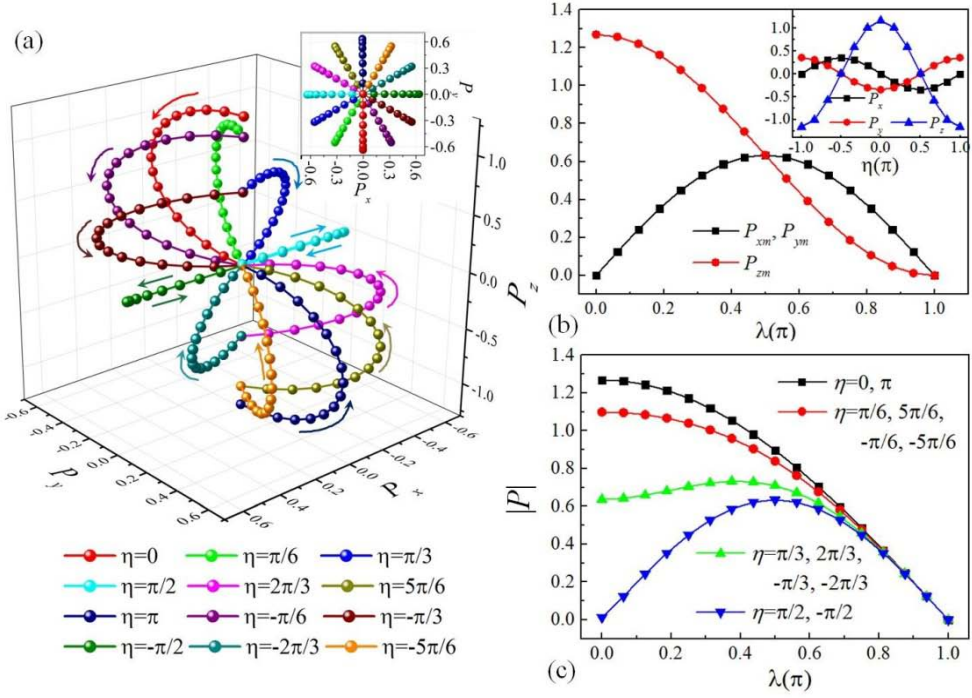

Figure 5 The electric polarization of SK crystal of $N_{sk}=-1$ with different $\lambda$ and $\eta$ obtained by simulation at $k=0$. (a) 3D chart of **P** as function of $\lambda$ with different $\eta$. The arrows show the variation direction of $\lambda$ from 0 to $\pi$. The inset presents the top view along $z$-axis. (b) $P_{xm}$, $P_{ym}$ and $P_{zm}$ as functions of $\lambda$. The inset presents the dependences of $P_x$, $P_y$ and $P_z$ on $\eta$ with $\lambda=0.1875\pi$. (c) |**P**| as a function of $\lambda$ for different $\eta$.

In addition, the isolated SK with $|N_{sk}|=2$ appears spontaneously as metastable state in the frustrated system [45]. The simulation indicates that its **V** can also be effectively controlled by **h** under topological protection, and tetra-meron state can be observed at $\lambda=0.5\pi$, as long as the anisotropy is not very strong. However, the helicity can not be controlled by electric field due to **P**=0 (See Note 3 of Supplemental Material for details [71]).

## IV. DYNAMIC EFFECTS OF VORTICITY MODULATION

The modulation of vector vorticity will generate some distinctive dynamic effects, which enrich the behaviors of SK, and provide more tunability. When a rotating **h** is applied on an isolated SK, a small linear shift could be observed, and its velocity depends on the frequency of $\boldsymbol{h}$ rotation. As demonstrated in Fig. 6(a), the trajectories of SK center (See Note 4 of Supplemental Material for details [71]) show the same length for the same angle of $\boldsymbol{h}$ rotation, but the orientation is associated with helicity and $N_{sk}$. It is seen that the orientation of trajectory depends on $\eta$ linearly. If $\eta$ is fixed, the trajectories of $N_{sk}$=1 and -1 present the opposite deviation from –x direction. The further simulation indicates that the straight motion is associated with the damping constant $\alpha$. For a $2\pi$ rotation of $\boldsymbol{h}$, the shift distance is denoted by $D_{hr}$, and the deviative angle from –x is $A_{hr}$. Both $D_{hr}$ and $A_{hr}$ increase with $\alpha$, which is different from the motion of antiferromagnetic domain wall driven by rotating magnetic field with the velocity independent of the damping $\alpha$ [77].

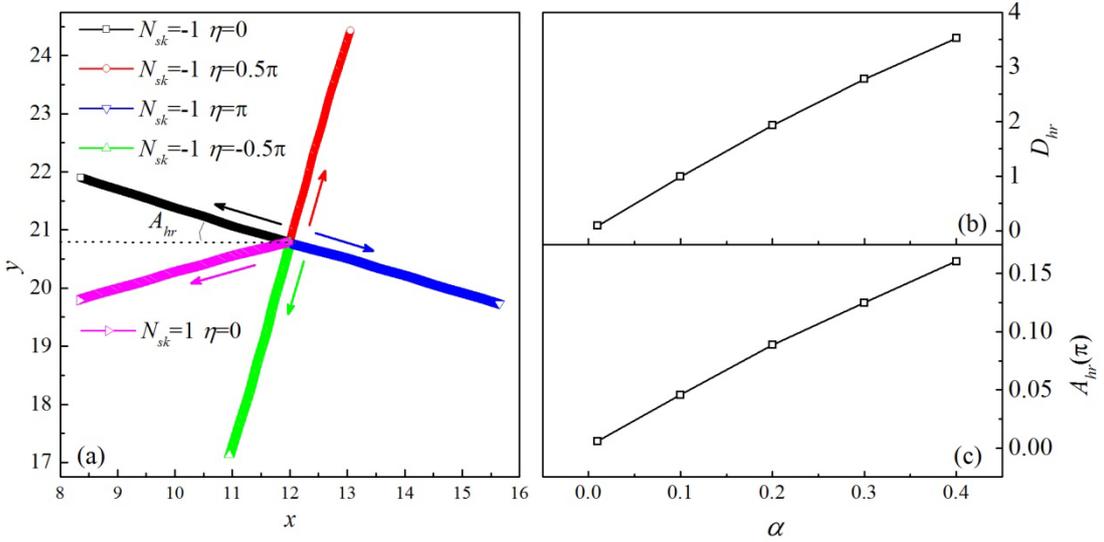

Figure 6 The linear motion of SK under $\boldsymbol{h}$ rotating around +x direction. (a) Trajectories of SK center under rotating $\boldsymbol{h}$ with $N_{sk}$= ±1 and different $\eta$. The arrow denotes the motion direction. (b) $D_{hr}$ and (c) $A_{hr}$ as functions of the damping $\alpha$ for $\eta$=0 and $N_{sk}$=-1.

Since the experiments revealed that the skyrmion in the metallic system can be driven and controlled by the ultralow electric currents [21,22], the spin polarized current induced dynamics of SK has attracted considerable interest due to the great potential for technical applications. The continuous modulation of vector vorticity will influence the SK dynamics driven by the spin polarized current obviously. For the spin transfer torque associated with the in-plane injection geometry of spin polarized current, no difference is observed for different vorticity, consistent with the previous report [60]. A more efficient way to drive skyrmion is the spin orbital torque resulting from the vertical injection of spin polarized current. Since a field-like part acts as a uniform field and its contribution to the SK dynamics is minor, only the damping-like term is considered for simplicity as follow.

$$\tau = -u\boldsymbol{S} \times (\boldsymbol{S} \times \boldsymbol{p}_{SC}) \tag{9}$$

where $p_{SC}$ is the direction of the spin current polarization vector along –y. $u$=0.02 is set in the reduced unit, representing the magnitude of spin torque. The SK dynamics driven by the spin polarized current is investigated with this additional torque term (Eq. 9) added to the LLG equation (Eq. 8). The vorticity is tuned by applying magnetic field $h$ rotating around +x, and the angle of $h$ from +z direction is denoted by $\beta$. When $\beta$=0, after a short transient regime, the SK moves along the circular path, as mentioned previously for the perpendicularly magnetized skyrmion [46,47]. Fig. 7(a) plots the trajectories produced by SKs with different $\beta$. From the same starting point, although initial motional orientations are different, they all stabilize into counterclockwise circles at last, except that the SK of $\beta$=0.5π remains stationary. When $h$ rotates from +z to –y, the diameter of the circular trajectory decreases significantly, and then increases symmetrically as $h$ rotates from –y to –z. Meanwhile, the helicity changes linearly with time and always couples to the circular motion of SK. Thus the variation of $\eta$ shows the same period (T) to that of circular motion, except $\beta$=0.5π. T exhibits the symmetrical dependence on $\beta$, as illustrated in Fig. 7(b). The SK moves at a constant velocity ($v_S$) along the circle, and $v_S$ also shows a symmetric dependence on $\beta$ with maximum at about $\beta$=0.04π and 0.96π. Therefore, the vorticity can be controlled by $h$ to modulate both period and velocity of the spin-polarized-current induced motion. In addition, it should be mentioned that the vorticity deviates from $h$ direction slightly due to the combination of magnetic field and spin torque, and this deviation even happens to $M$ of ferromagnetic state as shown in Fig. 7(c).

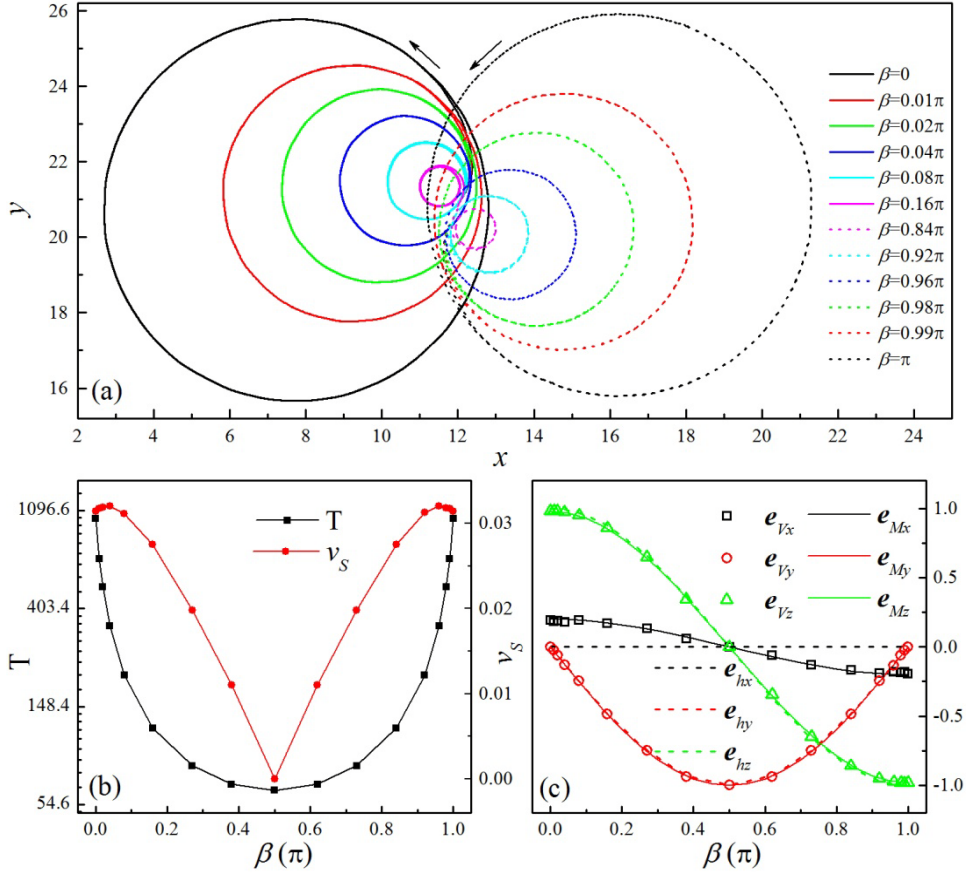

Figure 7 The vorticity modulation on the spin-polarized-current induced dynamics. (a) Trajectories of SKs driven by a damping-like spin torque under the magnetic field $\mathbf{h}$ with $\beta$ angle from +z direction. The arrow denotes the motion direction. (b) The period T and velocity $v_S$ of SK circular motion as functions of $\beta$, where Ln scale is used on the left to show the sharp variation of T. (c) The components of $\mathbf{e}_h$ ($e_{hx}$, $e_{hy}$, $e_{hz}$), $\mathbf{e}_M$ ($e_{Mx}$, $e_{My}$, $e_{Mz}$) and $\mathbf{e}_V$ ($e_{Vx}$, $e_{Vy}$, $e_{Vz}$) as functions of $\beta$, where $\mathbf{e}_V$ is obtained by SK simulation and $\mathbf{e}_M$ is calculated analytically for the ferromagnetic state.

## V. CONCLUSION

In summary, based on the theoretical analysis, we generalize the conventional scalar vorticity to a vector vorticity to describe the internal degree of freedom to full extent for SKs. The simplified vorticity angle, together with helicity angle, represents the whole rotation freedom of spin space. Rich morphologies of SKs with different vorticity and helicity angles are displayed, which can be transformed between each other by continuous deformation under topological protection. The centrosymmetric structure with frustration provides a perfect platform for realizing a controllable manipulation of these internal degrees of freedom. The simulation indicates that the vorticity can be tuned by magnetic field effectively for both SK crystal and isolated SK. The helicity can be further tuned by electric field in the case of

$|N_{sk}|$=1. Although the uniaxial anisotropy will induce distortion, the topological property can be well conserved and the effective modulation can also be realized, provided that the anisotropy is not very strong. Moreover, we discuss the dynamic effects of vorticity modulation for an isolated SK. The vorticity can be varied by rotating magnetic field to generate a straight motion, and it can also be tuned to control the spin-polarized-current induced dynamics.

**Acknowledgments**

We thank J. Chen for useful discussions. This work is supported by the research grants from the National Natural Science Foundation of China (Grant No. 11834002).